# Summarising Big Data: Common GitHub Dataset for Software Engineering Challenges

*Abdulkadir ŞEKER1,\**   *, Banu DİRİ2 , Halil ARSLAN1*

1*Sivas Cumhuriyet University, Department of Computer Engineering, Sivas/ TURKEY*

2*Yıldız Technical University, Department of Computer Engineering, İstanbul/ TURKEY*

**Abstract**

In open source software development environments; textual, numerical and relationship-based data generated are of interest to researchers. Various data sets are available for this data, which is frequently used in areas such as software engineering and natural language processing. However, since these data sets contain all the data in the environment, the problem arises in terabytes of data processing. For this reason, almost all of the studies using GitHub data use filtered data according to certain criteria. In this context, using a different data set in each study makes the comparison of the accuracy of the studies quite difficult. In order to solve this problem, a common dataset was created and shared with the researchers, which would allow to work on many software engineering problems.

## 1. Introduction

One of the most common among cloud-based open-source versioning systems is GitHub. GitHub can be thought of as a folder where we store more than 40 million repos[1].Has become the world's largest code server, hosted by more than 100 million developers[2].

On GitHub-like platforms, development processes are distributed. Developers can make different contributions to projects from any location, with comments, code additions, and topics that can fix bugs. In this way, large amounts of data are generated for researchers working on natural language processing and software engineering. In addition to these text data, the social relationships of users with each other and with projects (repo) also produce different content. Thanks to these features, GitHub data is of great importance and interest for academic and commercial studies.

In studies with GitHub data, researchers obtain data via the GitHub API or use some datasets. The most widely used of these datasets is known as GHTorrent [1]–[3]. The GHTorrent dataset was developed in the software engineering department of the Dutch Delft University [4]. The dataset presents all the data on the platform with the GitHub API, providing all information about repos and users.

The sizes of the GitHub datasets reach very high levels as they contain information from the entire platform. Working with data in such sizes is a problem in itself. When studies using GitHub data are examined, it is seen that almost all of them have created specific sub-datasets. Researchers have used the data in their studies by filtering the data according to their problems and possibilities. It is not possible to compare the successes of even the studies on similar subjects since each researcher works on the dataset that he/she has created.

To address this problem in the literature, the copy of the GHTorrent dataset containing GitHub data up to 2015 was localized and filtered with certain parameters, duplicate data was extracted, new fields to link the data were added, and a MongoDB dataset was created. In this way, a data set, which is much smaller than the GitHub environment, has been produced and shared, which will offer the opportunity to work on the challenges of software engineering in many areas such as task, user, project, and software development. It is possible to use this dataset easily during the algorithm or model development phase. The dataset was shared as a MongoDB archive for the developed codes to be applied to big data without being changed.

---

[1] Repository of projects
[2] https://en.wikipedia.org/wiki/gitHub

## 2. Definition of Problem

In studies using GitHub data, filtered datasets are used. Even if the problems studied are the same, since the datasets are specific to each study, the accuracy of the claimed success rates has become controversial. For example, the works are given in Table 1 are all about pull request (PR). However, as can be seen, they have used different types of content and filtered data of different types. In order to provide a solution to this situation and to be able to conduct rapid experiments at the beginning of the studies, a small dataset was designed and produced to represent the big data.

**Table 1.** The Filtered Dataset from some studies

| Study | Number of repo | Type of filter |
|---|---|---|
| Zhang-2014 [5] | 3587 | Has more than 100 PR |
| Veen-2015 [6] | 475 | - |
| Yu-2015 [7] | 40 | Has more than 100 PR and non-forked |
| Junior-2018 [8] | 32 | Has more than 100 PR and at least 5 developers |
| Zhao-2019 [9] | 74 | Develop with Java |

GHTorrent consists of a lot of information about projects, definitions and contents of the problem (*issue*), comments, information about pull requests, code commit activities, repos that users follow, etc. (Table 2).

**Table 2.** The Information of GHTorrent Dataset

| Domain | Collections |
|---|---|
| User | Users, Followers, Watchers, Repo Collaborators |
| Project | Projects, Forks |
| Development | Commits, CommitComments |
| Contribution | Issues, IssueComments, Pull Requests, Pull Request Comments |

This dataset offers GitHub data in different formats as SQL tables and MongoDB collections. In this study, a public subset of data was created with the version covering data up to 2015. With the MongoDB queries, some errors detected in the GHTorrent dataset have been corrected, free from repetitive data, link features that are noticed to be missing from the collections have been added, convert to a simple and easy-to-operate size and content.

While creating the dataset, all the data (approximately 750 GB) stored as MongoDB collections were downloaded to the local environment and put through the processes mentioned above. These processes have been developed in MongoDB in order to suitable for big data. The queries used for all transactions are shared on GitHub[3]. *Users* is selected as the restrictive collection when filtering the dataset. A dataset has been created with all the information associated with the 100 selected users. All collections have been shared as CSV format, which is the most suitable in terms of size.

## 3. Pre-Processes

### 3.1. Filtering

The 100 most active users on GitHub from the *Users* collection were selected[4]. Based on these users, all data that has any relationship with them was filtered from other collections. The steps for filtering are shown in detail in Figure 1.

---
[3] https://github.com/***/***
[4] https://gist.github.com/paulmillr/2657075

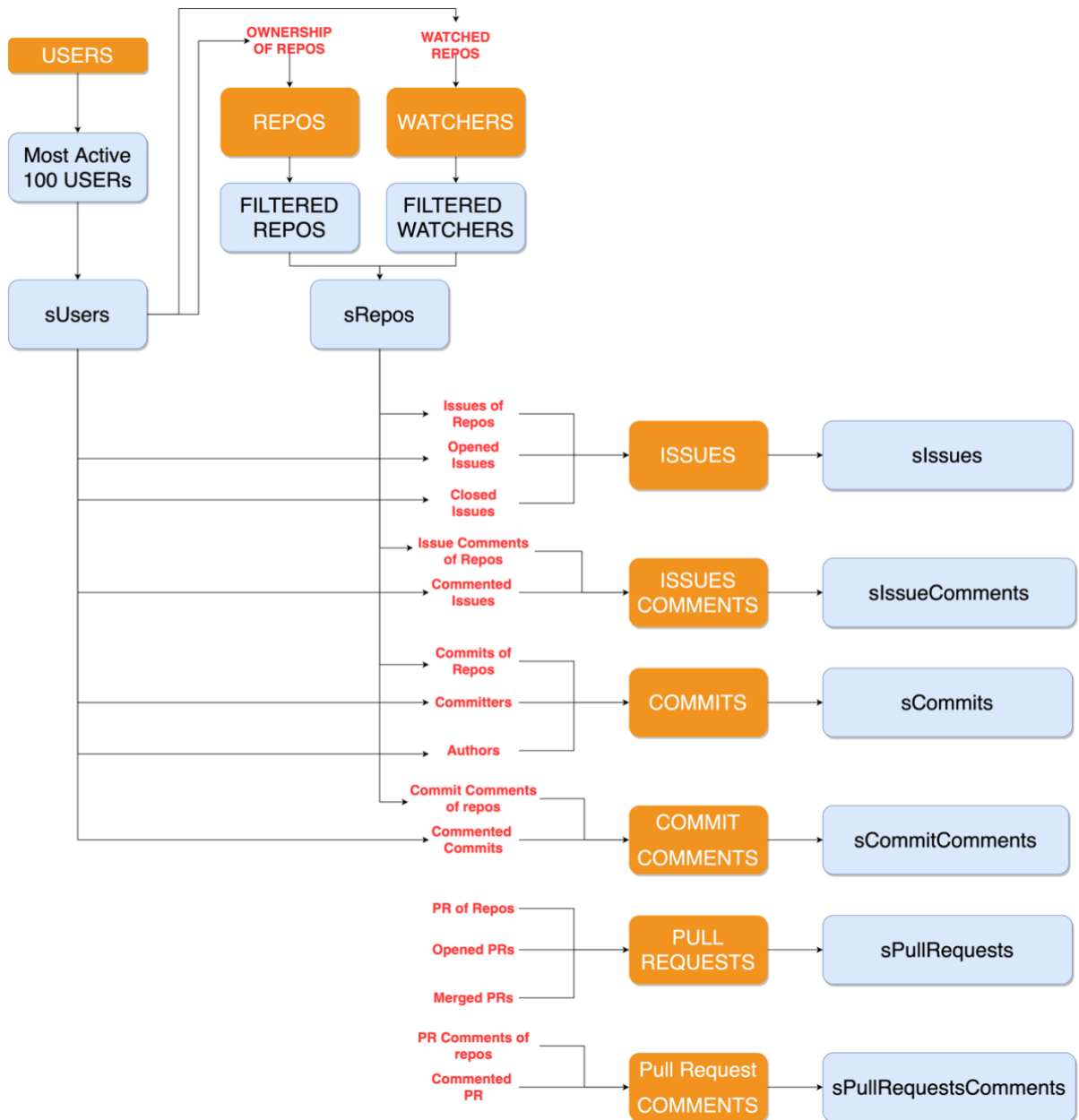

**Figure 1**. The filtering process with the most active 100 users

### 3.2. Solving mistaken or missed data

In order to obtain the dataset free from errors, it must be put through some process. In this section, the related MongoDB queries are given below each problem. Since the data selected in the dataset is planned in the user base, It is needed to determine the primary keys for collections. In this context, as a primary key (or distinctive feature); "*user id*" will be used on the data related to users, similarly *"repo id"* is used on the data associated with repos. Therefore, the existing problems with these fields must be eliminated.

1. Firstly, the documents[5] that has *null* value of these key features have been removed.

    ```
    db.Pull_requests.deleteMany(
    {user_id:null}
    );
    ```

---

[5] Document on mongodb platforms equals to row on SQL plaform)

2. It has been noticed that the dataset contains duplicate documents that are thought created by users or retrieve mistakenly. Documents containing repeated data in key fields such as *"user id"* or *"repo id"* or *"full name"* were also removed.

    ```
    db.Issues.aggregate([
    {$match: {"issue_id": {$nin:[null]}}},
    {$group: {  "_id": "$issue_id",
                "doc" : {"$first": "$$ROOT"}}},
    {$replaceRoot: { "newRoot": "$doc"}},
    {$out: "Issues"}

    ],

    {allowDiskUse:true}

    );
    ```

3. In some collections, it has been observed that the key fields do not exist or that their substitutes are not sufficient[6].
    a. *Issues*, *Watchers* collections have the *"owner"* and *"name"* fields of the repo. The *"full name"* field is created by combining two fields directly.

        ```
        db.Watchers.aggregate([
            {
             $addFields:
                { full_name:
                    { $concat: [ "$owner", "/", "$repo" ] }
                }
            },
            { $out: "Watchers"  }
        ]);
        ```

    b. In the *CommentCommits* collection, there is no *"owner"* and *"name"* fields. To obtain them firstly has been parsed the *"repo url"* field and extracted these fields. Then *"full name"* has been created as combining them. ( A sample *repo url*; *https://github.com/johndue/projectX*)

        ```
        db.CommentCommits.aggregate([
            {
             $addFields:
                { full_name:{ $concat:
                  [
                    {$arrayElemAt:[{$split:["$url", "/"]}, 4]},
                    "/",
                    {$arrayElemAt:[{$split:["$url", "/"]}, 5]}
                  ]
            }}},
            {$out:"CommentCommits"}
        ]);
        ```

---

[6] For example, in some collections, only the name is given as repo information. However, since the logic of fork is in Github, only repo name is not enough as a distinguishing feature. There may be different forked repos with the same repo name. For this reason, "full name" should be used as the single link field to add the \ textit {"repo id"} key field. This area consists of a combination of repo owner and repo name. These areas are given directly in some collections, while others are obtained with parsing some field.

c.  After the process *a* or *b*, the field of *"repo id"* has added to related collections as aggregate (join) with *Repos* collection on *"full name"*. The query below means; find the documents from both collections which have the same *"full name"*, then get the *"repo id"* of this document from a collection and add it to a related document in the other collection.

```
db.CommitComments.aggregate([
    {$lookup: { from: "repos",
                let: { item: "$full_name" },
                pipeline: [
                    { $match:
                        { $expr:
                            { $eq: ["$full_name","$$item"] }
                        }
                    },
                    { $project: { "repo_id": 1 } }
                ],
                as: "fromComments"}
    },
    {$replaceRoot:
      { newRoot:
        { $mergeObjects: [
            {$arrayElemAt:["$fromComments",0]},"$$ROOT"]}
        }
    },
    {$project: { 'fromComments': 0 } },
    {$out: "FilteredCommitComments"}
])
```

d.  Similarly, since there is no field of *"user id"* in the *Followers* collection, this field was added by join with *Users* collection on *login* field. (login is the name of user in GitHub database.)

After handling these adding key or link fields process, the sub dataset has been created. The comparison of size between the GHTorrent and proposed filtered data set are given in Table 3. As can be seen from Table 3, working with huge data causes serious time losses, especially during the algorithm or model development phase. In this context, it is thought that the proposed filtered data set will provide researchers with a common data pool but will also save time in their studies.

**Table 3.** The comparison of GHTorrent and proposed dataset

| Name of Collection | ~ Disk Size | | ~ Number of documents | |
| --- | --- | --- | --- | --- |
| | GHTorrent | Proposed | GHTorrent | Proposed |
| Users | 1 GB | 0.8 MB | 5,000,000 | 100 |
| Repos | 30 GB | 45 MB | 20,000,000 | 40,000 |
| Commits | 298 GB | 4 GB | 41,000,000 | 500,000 |
| Commit Comments | 1 GB | 90 MB | 2,000,000 | 250,000 |
| Issues | 11 GB | 3 GB | 17,000,000 | 3,000,000 |
| Issue Comments | 15 GB | 4 GB | 31,000,000 | 9,000,000 |
| Pull Requests | 23 GB | 5 GB | 8,000,000 | 1,500,000 |
| Pull Requests Comments | 6 GB | 1 GB | 5,000,000 | 1,200,000 |
| Followers | 1 GB | 20 MB | 7,000,000 | 130,000 |
| Forks | 8 GB | 5 MB | 11,000,000 | 7,000 |
| Watchers | 7 GB | 8 MB | 38,000,000 | 50,000 |
| RepoCollabrators | 1 GB | 2 MB | 5,000,000 | 4,000 |

## 3. Results and Discussion

In this study, a common dataset was proposed to the researchers working for the solution of software engineering challenges as a result of filtering with the GHTorrent dataset that contains an up-to-date copy of GitHub data. All filtering operations were performed in accordance with the big data in MongoDB environment so that the current versions of the data set can be extracted and extensions/enhancements can be made.

In the proposed dataset, the fields that will link between the collections have been added, the missing or repeated data noticed in the dataset have been removed. As a result, a clear and easy-to-operate GitHub dataset has been generated. The GitHub link given in the study can be used for download the dataset.

It is planned to studies on the software engineering challenges with this proposed dataset. Since this dataset is public, other researchers will be provided to compare their own studies in a clear and accurate manner.